\documentclass[11pt]{article}
\usepackage[utf8]{inputenc}
\usepackage[T1]{fontenc}
\usepackage{graphicx}
\usepackage{grffile}
\usepackage{longtable}
\usepackage{wrapfig}
\usepackage{rotating}
\usepackage[normalem]{ulem}
\usepackage{amsmath}
\usepackage{textcomp}
\usepackage{amssymb}
\usepackage{capt-of}
\usepackage{tikz}
\usepackage{xcolor}
\usepackage{cite}
\usepackage[cm]{fullpage}
\usepackage{mathtools,amsmath,amssymb,amsfonts,dsfont,mathrsfs,amsthm,latexsym}

\usepackage{bm}
\usepackage{centernot}
\usepackage{siunitx}
\usepackage{array}
\newcolumntype{L}[1]{>{\raggedright\let\newline\\\arraybackslash\hspace{0pt}}m{#1}}
\newcolumntype{C}[1]{>{\centering\let\newline\\\arraybackslash\hspace{0pt}}m{#1}}
\newcolumntype{R}[1]{>{\raggedleft\let\newline\\\arraybackslash\hspace{0pt}}m{#1}}
\usepackage{braket}
\usepackage{xparse}
\usepackage{breqn}

\DeclareDocumentCommand{\Ag}{ s o }{ \IfBooleanTF{#1}
    { \IfValueTF{#2}{ \bm{\mathcal{A}}_{(#2)} }{ \bm{\mathcal{A}} } }
    { \IfValueTF{#2}{    {\mathcal{A}}_{(#2)} }{    {\mathcal{A}} } } }
\DeclareDocumentCommand{\Af}{ s o }{ \IfBooleanTF{#1}
    { \IfValueTF{#2}{ \boldsymbol{A}_{(#2)} }{ \boldsymbol{A} } }
    { \IfValueTF{#2}{            {A}_{(#2)} }{            {A} } } }

\newcommand{\cdf}[1][]{{\boldsymbol{\mathcal{D}}}{#1}}

\DeclareDocumentCommand{\Fg}{ s o }{ \IfBooleanTF{#1}
    { \IfValueTF{#2}{ \bm{\mathcal{F}}_{(#2)} }{ \bm{\mathcal{F}} } }
    { \IfValueTF{#2}{    {\mathcal{F}}_{(#2)} }{    {\mathcal{F}} } } }
\DeclareDocumentCommand{\Ff}{ s o }{ \IfBooleanTF{#1}
    { \IfValueTF{#2}{ \boldsymbol{F}_{(#2)} }{ \boldsymbol{F} } }
    { \IfValueTF{#2}{            {F}_{(#2)} }{            {F} } } }

\DeclareMathOperator{\st}{\star}

\newcommand{\Mi}{\mathcal{M}}

\DeclareDocumentCommand{\PB}{ O{m} O{q} O{p} m m }{ \frac{ \partial #4 }{\partial {#2}^{#1} } \frac{ \partial #5 }{\partial {#3}_{#1} } - \frac{ \partial #4 }{\partial {#3}_{#1} } \frac{ \partial #5 }{\partial {#2}^{#1} } }

\DeclareDocumentCommand\Te{o o m }{\mathcal{T}{}^{#1}_{#2}(#3)}

\DeclareDocumentCommand{\BM}{ s }{ \IfBooleanTF{#1} {\hat{\bm{M}}}{\bm{M}} }
\DeclareDocumentCommand{\BN}{ s }{ \IfBooleanTF{#1} {\hat{\bm{N}}}{\bm{N}} }
\DeclareDocumentCommand{\BP}{ s }{ \IfBooleanTF{#1} {\hat{\bm{P}}}{\bm{P}} }
\DeclareDocumentCommand{\BQ}{ s }{ \IfBooleanTF{#1} {\hat{\bm{Q}}}{\bm{Q}} }
\DeclareDocumentCommand{\BR}{ s }{ \IfBooleanTF{#1} {\hat{\bm{R}}}{\bm{R}} }
\DeclareDocumentCommand{\BS}{ s }{ \IfBooleanTF{#1} {\hat{\bm{S}}}{\bm{S}} }
\DeclareDocumentCommand{\BU}{ s }{ \IfBooleanTF{#1} {\hat{\bm{U}}}{\bm{U}} }
\DeclareDocumentCommand{\BV}{ s }{ \IfBooleanTF{#1} {\hat{\bm{V}}}{\bm{V}} }

\NewDocumentCommand\MyAc{ m }{#1}
\DeclareDocumentCommand{\vif}{ t. t, t- s s m }{
  \RenewDocumentCommand\MyAc{ m }{##1}
  \IfBooleanT{#1}{\RenewDocumentCommand\MyAc{ m }{ \mathring{##1} } }
  \IfBooleanT{#2}{\RenewDocumentCommand\MyAc{ m }{ \tilde{##1} } }
  \IfBooleanT{#3}{\RenewDocumentCommand\MyAc{ m }{ \bar{##1} } }
  \IfBooleanTF{#4}
  { \IfBooleanTF{#5} { \hat{\MyAc{\boldsymbol{e}}}^{\hat{#6}} }{ \hat{\MyAc{\boldsymbol{e}}}^{{#6}} } }
  { \MyAc{\boldsymbol{e}}^{{#6}} } }
\DeclareDocumentCommand{\vi}{ t. t, t- s s m m}{
  \RenewDocumentCommand\MyAc{ m }{##1}
  \IfBooleanT{#1}{\RenewDocumentCommand\MyAc{ m }{ \mathring{##1} } }
  \IfBooleanT{#2}{\RenewDocumentCommand\MyAc{ m }{ \tilde{##1} } }
  \IfBooleanT{#3}{\RenewDocumentCommand\MyAc{ m }{ \bar{##1} } }
  \IfBooleanTF{#4}
  { \IfBooleanTF{#5} { \hat{\MyAc{e}}^{\hat{#6}}_{\hat{#7}} }{ \hat{\MyAc{e}}^{#6}_{{#7}} } }
  { \MyAc{e}^{{#6}}_{{#7}} } }

\DeclareDocumentCommand{\bt}{ t. t, t- s s m m m }{
  \RenewDocumentCommand\MyAc{ m }{##1}
  \IfBooleanT{#1}{\RenewDocumentCommand\MyAc{ m }{ \mathring{##1} } }
  \IfBooleanT{#2}{\RenewDocumentCommand\MyAc{ m }{ \tilde{##1} } }
  \IfBooleanT{#3}{\RenewDocumentCommand\MyAc{ m }{ \bar{##1} } }
  \IfBooleanTF{#4}
  { \IfBooleanTF{#5} { \hat{\MyAc{B}}_{{#6}}{}^{\hat{#7}}{}_{\hat{#8}} }{ \hat{\MyAc{\Gamma}}_{{#6}}{}^{{#7}}{}_{{#8}} } }
  { \MyAc{B}_{{#6}}{}^{{#7}}{}_{{#8}} } }

\DeclareDocumentCommand{\ct}{ t. t, t- s s m m m }{
  \RenewDocumentCommand\MyAc{ m }{##1}
  \IfBooleanT{#1}{\RenewDocumentCommand\MyAc{ m }{ \mathring{##1} } }
  \IfBooleanT{#2}{\RenewDocumentCommand\MyAc{ m }{ \tilde{##1} } }
  \IfBooleanT{#3}{\RenewDocumentCommand\MyAc{ m }{ \bar{##1} } }
  \IfBooleanTF{#4}
  { \IfBooleanTF{#5} { \hat{\MyAc{\Gamma}}_{{#6}}{}^{\hat{#7}}{}_{\hat{#8}} }{ \hat{\MyAc{\Gamma}}_{{#6}}{}^{{#7}}{}_{{#8}} } }
  { \MyAc{\Gamma}_{{#6}}{}^{{#7}}{}_{{#8}} } }
\DeclareDocumentCommand{\spif}{ t. t, t- s s m m }{
  \RenewDocumentCommand\MyAc{ m }{##1}
  \IfBooleanT{#1}{\RenewDocumentCommand\MyAc{ m }{ \mathring{##1} } }
  \IfBooleanT{#2}{\RenewDocumentCommand\MyAc{ m }{ \tilde{##1} } }
  \IfBooleanT{#3}{\RenewDocumentCommand\MyAc{ m }{ \bar{##1} } }
  \IfBooleanTF{#4}
  { \IfBooleanTF{#5} { \hat{\MyAc{\boldsymbol{\omega}}}^{\hat{#6}}{}_{\hat{#7}} }{ \hat{\MyAc{\boldsymbol{\omega}}}^{{#6}}{}_{{#7}} } }
  { \MyAc{\boldsymbol{\omega}}^{{#6}}{}_{{#7}} } }
\DeclareDocumentCommand{\spi}{ t. t, t- s s m m m }{
  \RenewDocumentCommand\MyAc{ m }{##1}
  \IfBooleanT{#1}{\RenewDocumentCommand\MyAc{ m }{ \mathring{##1} } }
  \IfBooleanT{#2}{\RenewDocumentCommand\MyAc{ m }{ \tilde{##1} } }
  \IfBooleanT{#3}{\RenewDocumentCommand\MyAc{ m }{ \bar{##1} } }
  \IfBooleanTF{#4}
  { \IfBooleanTF{#5} { \hat{\MyAc{{\omega}}}_{\hat{#6}}{}^{\hat{#7}}{}_{\hat{#8}} }{ \hat{\MyAc{{\omega}}}_{{#6}}{}^{{#7}}{}_{{#8}} } }
  { \MyAc{{\omega}}_{{#6}}{}^{{#7}}{}_{{#8}} } }

\DeclareDocumentCommand{\rif}{ t. t, t- s s m m }{
  \RenewDocumentCommand\MyAc{ m }{##1}
  \IfBooleanT{#1}{\RenewDocumentCommand\MyAc{ m }{ \mathring{##1} } }
  \IfBooleanT{#2}{\RenewDocumentCommand\MyAc{ m }{ \tilde{##1} } }
  \IfBooleanT{#3}{\RenewDocumentCommand\MyAc{ m }{ \bar{##1} } }
  \IfBooleanTF{#4}
  { \IfBooleanTF{#5} { \hat{\MyAc{\bm{\mathcal{R}}}}{}^{\hat{#6}}{}_{\hat{#7}} }{ \hat{\MyAc{\bm{\mathcal{R}}}}{}^{{#6}}{}_{{#7}} } }
  { \MyAc{\bm{\mathcal{R}}}{}^{{#6}}{}_{{#7}} } }
\DeclareDocumentCommand{\ri}{ t. t, t- s s m m m }{
  \RenewDocumentCommand\MyAc{ m }{##1}
  \IfBooleanT{#1}{\RenewDocumentCommand\MyAc{ m }{ \mathring{##1} } }
  \IfBooleanT{#2}{\RenewDocumentCommand\MyAc{ m }{ \tilde{##1} } }
  \IfBooleanT{#3}{\RenewDocumentCommand\MyAc{ m }{ \bar{##1} } }
  \IfBooleanTF{#4}
  { \IfBooleanTF{#5} { \hat{\MyAc{\mathcal{R}}}_{{#6}}{}^{\hat{#7}}{}_{\hat{#8}} }{ \hat{\MyAc{\mathcal{R}}}_{{#6}}{}^{{#7}}{}_{{#8}} } }
  { \MyAc{\mathcal{R}}_{{#6}}{}^{{#7}}{}_{{#8}} } }

\newcommand{\Rif}[2]{\bm{\mathcal{R}}^{{#1}}{}_{{#2}}}

\DeclareDocumentCommand{\kf}{ t. t, t- s s m m }{
  \RenewDocumentCommand\MyAc{ m }{##1}
  \IfBooleanT{#1}{\RenewDocumentCommand\MyAc{ m }{ \mathring{##1} } }
  \IfBooleanT{#2}{\RenewDocumentCommand\MyAc{ m }{ \tilde{##1} } }
  \IfBooleanT{#3}{\RenewDocumentCommand\MyAc{ m }{ \bar{##1} } }
  \IfBooleanTF{#4}
  { \IfBooleanTF{#5} { \hat{\MyAc{\bm{\mathcal{K}}}}^{\hat{#6}}{}_{\hat{#7}} }{ \hat{\MyAc{\bm{\mathcal{K}}}}^{{#6}}{}_{{#7}} } }
  { \MyAc{\bm{\mathcal{K}}}^{{#6}}{}_{{#7}} } }
\DeclareDocumentCommand{\ko}{ t. t, t- s s m m m }{
  \RenewDocumentCommand\MyAc{ m }{##1}
  \IfBooleanT{#1}{\RenewDocumentCommand\MyAc{ m }{ \mathring{##1} } }
  \IfBooleanT{#2}{\RenewDocumentCommand\MyAc{ m }{ \tilde{##1} } }
  \IfBooleanT{#3}{\RenewDocumentCommand\MyAc{ m }{ \bar{##1} } }
  \IfBooleanTF{#4}
  { \IfBooleanTF{#5} { \hat{\MyAc{\mathcal{K}}}_{\hat{#6}}{}^{\hat{#7}}{}_{\hat{#8}} }{ \hat{\MyAc{\mathcal{K}}}_{{#6}}{}^{{#7}}{}_{{#8}} } }
  { \MyAc{\mathcal{K}}_{{#6}}{}^{{#7}}{}_{{#8}} } }

\DeclareDocumentCommand{\tf}{ t. t, t- s s m }{
  \RenewDocumentCommand\MyAc{ m }{##1}
  \IfBooleanT{#1}{\RenewDocumentCommand\MyAc{ m }{ \mathring{##1} } }
  \IfBooleanT{#2}{\RenewDocumentCommand\MyAc{ m }{ \tilde{##1} } }
  \IfBooleanT{#3}{\RenewDocumentCommand\MyAc{ m }{ \bar{##1} } }
  \IfBooleanTF{#4}
  { \IfBooleanTF{#5} { \hat{\MyAc{\bm{\mathcal{T}}}}^{\hat{#6}} }{ \hat{\MyAc{\bm{\mathcal{T}}}}^{{#6}} } }
  { \MyAc{\bm{\mathcal{T}}}^{{#6}} } }

\newcommand{\tors}[3]{\mathcal{T}{}_{#1}{}^{#2}{}_{#3}}

\newcommand{\comm}[2]{\left[#1,#2\right]}

\renewcommand{\set}[1]{\ensuremath{\Set{ #1 }}}

\newcommand{\Tr}{\operatorname{Tr}}

\newcommand{\beq}{\begin{equation}}
\newcommand{\eeq}{\end{equation}}
\newcommand{\ber}{\begin{eqnarray}}
\newcommand{\eer}{\end{eqnarray}}

\NewDocumentCommand{\tak}{ s m m}{
  \IfBooleanTF{#1}{ \big( {#2} \big) \big[ {#3} \big] }
              { \big( {#2} \big] \big[ {#3} \big) }
}

\newcommand{\dn}[2]{{\mathrm{d}}^{#1}\!{#2}\;}

\newcommand\UTFSM{Departamento de F\'isica, Universidad T\'{e}cnica Federico Santa Mar\'\i a, \\ Casilla 110-V, Valpara\'iso, Chile}

\newcommand\CCTVal{Centro Cient\'ifico Tecnol\'ogico de Valpara\'iso, \\ Casilla 110-V, Valpara\'\i so, Chile}

\usepackage{hyperref}
\hypersetup{linktocpage,colorlinks=true,allcolors=blue}
\allowdisplaybreaks[3]
\begin{document}

\title{Beyond Einstein: a polynomial affine model of gravity}

\author{Oscar
  {Castillo-Felisola}\thanks{o.castillo.felisola@gmail.com}\\
  \CCTVal\\and\\\UTFSM.}

\maketitle

\begin{abstract}
  We show that the effective field equations for a recently
  formulated polynomial affine model of gravity, in the sector of a
  torsion-free connection, accept general Einstein manifolds---with or
  without cosmological constant---as solutions. Moreover, the
  effective field equations are partially those obtained from a
  gravitational Yang--Mills theory known as
  Stephenson--Kilmister--Yang theory. Additionally, we find a
  generalization of a minimally coupled massless scalar field in
  General Relativity within a ``minimally'' coupled scalar field in
  this affine model. Finally, we present the road map to finding general
  solutions to the effective field equations with either isotropic and
  cosmologic (i.e. homogeneous and isotropic) symmetry.
\end{abstract}


\section{Introduction}
\label{sec.intro}
During the last century (approximately), we reached a high level of
understanding of the four fundamental interactions, i.e.,
Electromagnetic, Weak, Strong and Gravitational. However, our
understanding splits into two streams: The first describes three kinds
of interactions and includes them into a single model, called standard
model of particle physics; while the second covers only the
gravitational interactions.

The interactions within the standard model of particles are described
by connection fields, modelled by gauge theories, and their
quantization procedure is successfully applied. On the other hand, the
gravitational interaction, as formulated by Einstein
\cite{einstein1915allgemeinen} and Hilbert \cite{hilbert1915grundlagen},
is described by the metric field, whose model does not fit into the
category of gauge theory, and its quantization procedure is not yet
well-defined
\cite{tHooft:1973us,tHooft:1974bx,Deser:1974cz,Deser:1974cy,DeWitt:1967yk,DeWitt:1967ub,DeWitt:1967uc} (see
Ref. \cite{Rovelli:2000aw} for a historical review).

The above suggests that some of the theoretical problems encountered
when trying to quantize the gravitational interactions,\footnote{The issue is not only related to quantization, but also to
generality. Mathematically, the only ingredient needed to define the
curvature is the connection. Thus, considering connections defined by
a more fundamental field is not the most general set up.} are due
to fact that it is formulated as a field theory for the metric, and
not as a theory for a connection. Therefore, there have been several
attempts of describing the gravitational interaction from an affine
view point, by using only the connection as fundamental field of the
model \cite{Eddington1923math,schrodinger1950space}, but these
descriptions were not very successful. In addition, Cartan's proposal of
considering the same field equations (or action) than
Einstein--Hilbert, but with more general connections, see Refs.
\cite{Cartan1922,Cartan1923,Cartan1924,Cartan1925}, was left aside
because the field equations, in pure gravity, impose the vanishing
torsion.

After the works by Kibble \cite{Kibble:1961ba} and Sciama
\cite{Sciama:1964wt}, it was understood that once gravity couples to
matter could exist a nontrivial torsion compatible with the setup, and
the search of viable affine
\cite{Krasnov:2007ei,Plebanski:1977zz,kijowski78_new_variat_princ_gener_relat,ferraris81_gener_relat_is_gauge_type_theor,ferraris82_equiv_relat_theor_gravit,Krasnov:2007uu,Krasnov:2007ky,Krasnov:2011pp,poplawski07_on_the_nonsy_purel_affin_gravit,Poplawski:2012bw,Skirzewski:2014eta,Castillo-Felisola:2015cqa}
(and metric-affine \cite{Hehl:1994ue}) models of gravity become relevant
again.

Before continuing our arguments for the \emph{necessity} of considering
affine generalisations of General Relativity, we shall briefly remind
some basic concepts in geometry. In the antiquity, the plane geometry
was built essentially with the aid of a (straight) rule and a
compass. Counter-intuitively, the compass was used to measure
distances, while the rule was used to define parallelism.  In modern
differential geometry language, the object that allows us to measure
distances is the metric, (\(g_{\mu\nu}\)), while the one associated
with the concept of parallelism is the connection
(\(\ct{}{\lambda}{\mu\nu}\)). 

Although the concepts of distance and parallelism are independent,
there exists a (unique) particular case in which the concepts relate
with each other, and thus one needs just of the metric---while the
connection is a potential for the metric. This particular case is
known as Riemannian geometry, and General Relativity stands on such
particular construction.

The geometries with general connection are called with the adjective
affine, and are characterised by its curvature
(\(\ri{\mu\nu}{\lambda}{\rho}\)), torsion (\(\tors{}{\lambda}{\mu\nu}
= \ct{}{\lambda}{\mu\nu} - \ct{}{\lambda}{\nu\mu}\)) and the metricity
condition (\(\mathcal{Q}_{\lambda \mu\nu} = \nabla^\Gamma_\lambda
g_{\mu\nu}\)) \cite{Hehl:1994ue,schouten2013ricci,ortin15_gravit_strin}.
Notice that in Riemannian geometries, either
\(\tors{}{\lambda}{\mu\nu}\) and \(\mathcal{Q}_{\lambda \mu\nu}\)
vanish, and the only quantity that characterises the manifolds is the
curvature.

An argument to consider affine geometries, to describe gravitational
interactions, is that it introduces new degrees of freedom into the
model (other than a spin two field), which might be interpreted as
matter---from the view point of General Relativity---contributing to
the dark sector of the Universe
\cite{Poplawski:2010jv,Poplawski:2011wj}, inflation
\cite{Poplawski:2010kb}, exotic cosmologies
\cite{castillo-felisola16_kaluz_klein_cosmol_from_five}, and diverse
particle physics effects \cite{Belyaev:1998ax,Belyaev:2007fn,Fabbri:2012yg,Fabbri:2012zd,Capozziello:2013dja,Castillo-Felisola:2013jva,Castillo-Felisola:2014iia,Castillo-Felisola:2014xba,Castillo-Felisola:2015ema}.

In the following sections we will analyse a recently proposed model
called \emph{Polynomial Affine Gravity}
\cite{Skirzewski:2014eta,Castillo-Felisola:2015cqa}, which is built up
with an affine connection as sole field, and under the premise of
preserving the whole group of diffeomorphisms.

\section{Polynomial affine gravity: the model}
\label{sec.PAG}
The connection is the field that allows us to define the notion of
parallelism as follows. Given a connection
\(\hat{\Gamma}^\mu{}_{\rho\sigma},\) one defines a \emph{covariant}
derivative, (\(\nabla^{\hat{\Gamma}}\)), such that, if the directional
derivative of a geometrical object, \(\mathcal{V}\), along a vector (\(X\)) vanishes, one says
that the object is parallel transported along (the integral curve
defined by) the vector,
\begin{equation*}
  \nabla^{\hat{\Gamma}}_X \mathcal{V} = 0.
\end{equation*}

The affine connection accepts a decomposition on
irreducible components as
\begin{equation}
  \hat{\Gamma}^\mu{}_{\rho\sigma} = \hat{\Gamma}^\mu{}_{(\rho\sigma)} + \hat{\Gamma}^\mu{}_{[\rho\sigma]} = {\Gamma}^\mu{}_{\rho\sigma} + \epsilon_{\rho\sigma\lambda\kappa}T^{\mu,\lambda\kappa}+A_{[\rho}\delta^\mu_{\nu]},
\end{equation}
where \({\Gamma}^\mu{}_{\rho\sigma} =
\hat{\Gamma}^\mu{}_{(\rho\sigma)}\) is symmetric in the lower indices,
\(A_\mu\) is a vector field corresponding to the trace of torsion, and
\(T^{\mu,\lambda\kappa}\) is a Curtright-like field
\cite{Curtright:1980yk},\footnote{Notice that the Curtright-like field is defined as the
quasi-Hodge dual of the traceless part of the torsion.} satisfying \(T^{\kappa,\mu\nu } = -
T^{\kappa,\nu\mu }\) and
\(\epsilon_{\lambda\kappa\mu\nu}T^{\kappa,\mu\nu }=0\).\footnote{Since no metric is present, the epsilon symbols are not related
by raising (lowering) their indices, but instead we demand that
\(\epsilon^{\delta\eta\lambda\kappa} \epsilon_{\mu\nu\rho\sigma} = 4!
\delta^{\delta}{}_{[\mu}\delta^\eta{}_{\nu}\delta^{\lambda}{}_{\rho}
\delta^\kappa{}_{\sigma]}\).}

Using the above decomposition, we need to build the most general
action preserving diffeomorphisms. In order to guarantee the correct
transformation of the Lagrangian density, the geometrical objects used
to write down the action are a Curtright (\(T^{\mu,\nu\lambda}\)), a
vector (\(A_\mu\)), the covariant derivative defined with the
Levi-Civita connection (\(\nabla_{\mu}\)), both Levi-Civita tensors
(\(\epsilon_{\mu\nu\lambda\rho}\) and \(\epsilon^{\mu\nu\lambda\rho}\)),
and the Riemannian curvature (\(R_{\mu\nu}{}^\lambda{}_\rho\)).  Since
the Riemannian curvature is defined as the commutator of the covariant
derivative, it is not an independent field, so it will be left out of
the analysis, and only five ingredients remain. In
Ref. \cite{Castillo-Felisola:2015cqa}, a method of \emph{dimensional
analysis} was introduced to ensure that all possible terms were taken
into account, and the general action---up to boundary and topological
terms---is
\begin{dmath}
  \label{4dfull}
  S[{\Gamma},T,A] =
  \int\dn{4}{x}\Bigg[
    B_1\, R_{\mu\nu}{}^{\mu}{}_{\rho} T^{\nu,\alpha\beta}T^{\rho,\gamma\delta}\epsilon_{\alpha\beta\gamma\delta}
    +B_2\, R_{\mu\nu}{}^{\sigma}{}_\rho T^{\beta,\mu\nu}T^{\rho,\gamma\delta}\epsilon_{\sigma\beta\gamma\delta}
    +B_3\, R_{\mu\nu}{}^{\mu}{}_{\rho} T^{\nu,\rho\sigma}A_\sigma
    +B_4\, R_{\mu\nu}{}^{\sigma}{}_\rho T^{\rho,\mu\nu}A_\sigma
    +B_5\, R_{\mu\nu}{}^{\rho}{}_\rho T^{\sigma,\mu\nu}A_\sigma
    +C_1\, R_{\mu\rho}{}^{\mu}{}_\nu \nabla_\sigma T^{\nu,\rho\sigma}
    +C_2\, R_{\mu\nu}{}^{\rho}{}_\rho \nabla_\sigma T^{\sigma,\mu\nu} 
    +D_1\, T^{\alpha,\mu\nu}T^{\beta,\rho\sigma}\nabla_\gamma T^{(\lambda, \kappa) \gamma}\epsilon_{\beta\mu\nu\lambda}\epsilon_{\alpha\rho\sigma\kappa}
    +D_2\,T^{\alpha,\mu\nu}T^{\lambda,\beta\gamma}\nabla_\lambda T^{\delta,\rho\sigma}\epsilon_{\alpha\beta\gamma\delta}\epsilon_{\mu\nu\rho\sigma}
    +D_3\,T^{\mu,\alpha\beta}T^{\lambda,\nu\gamma}\nabla_\lambda T^{\delta,\rho\sigma}\epsilon_{\alpha\beta\gamma\delta}\epsilon_{\mu\nu\rho\sigma}
    +D_4\,T^{\lambda,\mu\nu}T^{\kappa,\rho\sigma}\nabla_{(\lambda} A_{\kappa)} \epsilon_{\mu\nu\rho\sigma}
    +D_5\,T^{\lambda,\mu\nu}\nabla_{[\lambda}T^{\kappa,\rho\sigma} A_{\kappa]} \epsilon_{\mu\nu\rho\sigma}
    +D_6\,T^{\lambda,\mu\nu}A_\nu\nabla_{(\lambda} A_{\mu)}
    +D_7\,T^{\lambda,\mu\nu}A_\lambda\nabla_{[\mu} A_{\nu]} 
    +E_1\,\nabla_{(\rho} T^{\rho,\mu\nu}\nabla_{\sigma)} T^{\sigma,\lambda\kappa}\epsilon_{\mu\nu\lambda\kappa}
    +E_2\,\nabla_{(\lambda} T^{\lambda,\mu\nu}\nabla_{\mu)} A_\nu
    +T^{\alpha,\beta\gamma}T^{\delta,\eta\kappa}T^{\lambda,\mu\nu}T^{\rho,\sigma\tau}
    \Big(F_1\,\epsilon_{\beta\gamma\eta\kappa}\epsilon_{\alpha\rho\mu\nu}\epsilon_{\delta\lambda\sigma\tau}
    +F_2\,\epsilon_{\beta\lambda\eta\kappa}\epsilon_{\gamma\rho\mu\nu}\epsilon_{\alpha\delta\sigma\tau}\Big) 
    +F_3\, T^{\rho,\alpha\beta}T^{\gamma,\mu\nu}T^{\lambda,\sigma\tau}A_\tau \epsilon_{\alpha\beta\gamma\lambda}\epsilon_{\mu\nu\rho\sigma}
    +F_4\,T^{\eta,\alpha\beta}T^{\kappa,\gamma\delta}A_\eta A_\kappa\epsilon_{\alpha\beta\gamma\delta}\Bigg].
\end{dmath}

Despite the complex structure of the action, it shows very interesting
features: (a) The structure is rigid (it does not accept extra terms),
which forbid the appearance of counter-terms, if one would like to
quantise the model; (b) All coupling constants are dimensionless,
which might be a hint of conformal invariance of the model; (c) The
action turns out to be power-counting renormalizable, which does not
guarantee renormalizability, but is a nice feature; (d) The structure
of the model yields no three-point graviton vertices, which might
allow to overcome the \emph{no-go} theorems found in
Refs. \cite{McGady:2013sga,Camanho:2014apa}.

\section{Limit of vanishing torsion}
\label{sec.torsionless}
We now want to restrict ourselves to the limit of vanishing torsion,
which simplify the comparison between our model and General
Relativity. the vanishing torsion limit---equivalent to take
\(T^{\lambda,\mu\nu} \to 0\) and \(A_\mu \to 0\)---cannot be taken at
the action level, but in the field equations, and the limit is a
consistent truncation of the whole field equations
\cite{Castillo-Felisola:2015cqa}.

The only nontrivial field equation after the limit will be the one
for the Curtright-like field, \(T^{\nu,\mu\rho}\),
\begin{equation}
  \nabla_{[\rho} R_{\mu]\nu} + \kappa \nabla_{\nu} R_{\mu\rho}{}^\lambda{}_\lambda = 0,
  \label{almostSimpleEOM}
\end{equation}
with \(\kappa\) a constant related with the original couplings of the
model. These field equations are simpler if one restricts to
connections compatible with a volume form, also known as equi-affine
\cite{schouten2013ricci,nomizu1994affine,MO-Bryant02}, which assures
that the Ricci tensor of the connection is symmetric, and the
contraction of the last indices vanishes, thus the equations are
\begin{equation}
  \nabla_{[\rho} R_{\mu]\nu} = 0.
  \label{SimpleEOM}
\end{equation}

Equation \eqref{SimpleEOM} is a generalization of Einstein's field
equation in vacuum. This can be seem as follows: All Einstein
manifolds posses Ricci tensor proportional to the metric, \(R_{\mu\nu}
\propto g_{\mu\nu}\), the metricity condition thus ensures that every
vacuum solution to the Einstein's equations solves the (simplified)
field equations of our model.

Moreover, the Eq. \eqref{SimpleEOM} is related through the second Bianchi
identity to the harmonic curvature
condition \cite{bourguignon1981varietes},
\begin{equation}
  \label{harm-curv}
  \nabla_\lambda R_{\mu\nu}{}^\lambda{}_\rho = 0.
\end{equation}

The Eqs. \eqref{SimpleEOM} and \eqref{harm-curv} accept a geometrical
interpretation equivalent to that of the field equations of a pure
Yang--Mills theory, which in the language of differential forms are
\begin{align}
  \cdf \Ff* &= 0, & \cdf \st \Ff* &= 0,
\end{align}
where \(\Ff* = \cdf \Af*\) is the field strength 2-form (the curvature
2-form of the connection in the principal bundle, see for example
Ref. \cite{bourguignon1982yang,Nakahara}), and the operator \(\st\)
denotes the Hodge star. Now, these Yang--Mills field equations are
obtained from the variation of the action functional
\begin{equation}
  S_{\textsc{ym}} = \int \Tr \Big( \Ff* \st \Ff* \Big),
\end{equation}
and the Jacobi identity for the covariant derivative.

Similarly, Eq. \eqref{SimpleEOM} (equivalently
Eq. \eqref{harm-curv}) can be obtained from an effective gravitational
Yang--Mills functional
action \cite{stephenson1958quadratic,kilmister1961use,Yang1974},
\begin{equation}
  \label{SKY}
  S_{\textsc{ym}} = \int \Tr \left( \Rif{}{} \st \Rif{}{} \right) = \int \left( \Rif{a}{b} \st \Rif{b}{a} \right),
\end{equation}
where \(\Rif{}{} \in \Omega^2(\Mi, T^*\Mi \otimes T\Mi)\) is the
curvature two-form, the operator \(\st\) denotes the Hodge star, and the
trace is taken on the bundle indices (see
Ref. \cite{bourguignon1982yang}).

The gravitational model described by the action in Eq. \eqref{SKY} is
called Stephenson--Kilmister--Yang (or SKY for short), and its
physical interpretation relies---as in General Relativity---in the
fact that the metric is the fundamental field for describing the
gravitational interaction. In that case, the field equations are third
order partial differential equations, and there are several
undesirable behaviours due to this characteristic of the
equations. However, in our model the field mediating the gravitational
interaction is the connection and, therefore, the Eqs.
\eqref{SimpleEOM} and  \eqref{harm-curv} are second order field equations
for the components of the connection.

It is worth noticing that, according to the arguments in
Refs. \cite{McGady:2013sga,Camanho:2014apa}, the SKY theory is not
renormalizable. However, it is possible that the polynomial affine
gravity could be renormalizable, in the sense that SKY is an effective
describtion for the torsionless limit of polynomial affine gravity.

\section{Polynomial affine gravity coupled to a scalar field}
\label{sec.scalar}
In the standard formulation of physical theories (even in flat
spacetimes), the metric is a required ingredient. The metric and its
inverse define an homomorphism between the tangent and cotangent
bundle, allowing to build the kinetic energy term in the action
\begin{equation*}
  \partial_\mu \phi \partial^\mu \phi = g^{\mu\nu} \partial_\mu \phi \partial_\nu \phi.
\end{equation*}
Therefore, the inclusion of matter within models with no necessity of
a metric, is a nontrivial task.

Inspired in the method of \emph{dimensional analysis} introduced in
Ref. \cite{Castillo-Felisola:2015cqa}, we attempt to couple a scalar
field to the polynomial affine gravity by defining the most general,
symmetric \(\binom{2}{0}\)-tensor density, \(\mathfrak{g}^{\mu\nu}\),
built with the available fields, and use it to build Lagrangian
densities for the matter content. It can be shown that such a density
is given by
\begin{equation}
  \mathfrak{g}^{\mu\nu} = \alpha \, \nabla_\lambda T^{\mu,\nu\lambda} + \beta \, A_\lambda T^{\mu,\nu\lambda} + \gamma \, \epsilon_{\lambda\kappa\rho\sigma} T^{\mu, \lambda\kappa} T^{\nu, \rho\sigma},
  \label{geng}
\end{equation}
with \(\alpha\), \(\beta\) and \(\gamma\) arbitrary coefficients.

Therefore, the action defined by the ``kinetic term'' is
\begin{equation}
  \label{ScalarAction}
  S_\phi = -  \int \dn{4}{x} \Big( \alpha \, \nabla_\lambda T^{\mu,\nu\lambda}  + \beta \, A_\lambda T^{\mu,\nu\lambda} + \gamma \, \epsilon_{\lambda\kappa\rho\sigma} T^{\mu, \lambda\kappa} T^{\nu, \rho\sigma} \Big) \partial_\mu\phi\partial_\nu\phi.
\end{equation}

Remarkably, it induces a nontrivial contribution to the field
equations once we restrict to the torsionless sector.  The field
nontrivial equations, when the scalar field is turned on, are\footnote{We have fixed the coefficient \(C_1 = 1\).} 
\begin{equation*}
  \nabla_{[\sigma} R_{\rho]\mu}{}^{\mu}{}_\nu - {C_2} \nabla_\nu  R_{\rho\sigma}{}^{\mu}{}_\mu - \alpha \nabla_{[\sigma} \Big( \partial_{\rho]}\phi \partial_\nu\phi \Big) = 0,
\end{equation*}
which for equi-affine connections simplifies to 
\begin{equation}
  \nabla_{[\sigma} R_{\rho]\nu} - \alpha \nabla_{[\sigma} \Big( \partial_{\rho]} \phi \partial_\nu \phi \Big) = 0.
  \label{SimpleEOMwS}
\end{equation}

Equation \eqref{SimpleEOMwS} can be integrated once, and the solution
takes the familiar form,
\begin{equation*}
  R_{\mu\nu} - \alpha \partial_{\mu} \phi \partial_{\nu} \phi = \Lambda g_{\mu\nu},
  \label{eq.scalar1}
\end{equation*}
where the integration (covariantly) constant tensor, which is
invertible and symmetric, has been suggestively denoted by \(\Lambda
g_{\mu\nu}\). The above equation can be written in the more
conventional form
\begin{equation}
  R_{\mu\nu} - \frac{1}{2} g_{\mu\nu} R + \Lambda g_{\mu\nu} = \alpha \Big( \partial_{\mu} \phi \partial_{\nu} \phi -
  \frac{1}{2} g_{\mu\nu} \big( \partial\phi \big)^2 \Big).
  \label{eq.scalar2}
\end{equation}
Thus, we have been able to recover a set of equations, similar to
those of General Relativity coupled with a free, massless scalar
field, but with an arbitrary rank two, symmetric, covariantly constant
tensor playing the role of a metric.

Notice that, from the action in Eq. \eqref{ScalarAction} it is not
possible to obtain the field equation for the scalar \(\phi\), when
the vanishing torsion limit is taken. Nonetheless, the second Bianchi
identity in Eq. \eqref{eq.scalar1} imposes
\begin{equation}
  \nabla^\mu \partial_{\mu} \phi = 0.
\end{equation}
This condition is, in the sense argued in
Ref. \cite{Bekenstein:2014uwa}, the equation of motion for the scalar
field.

\section{Finding symmetric ansätze}
\label{sec.symm}
The usual procedure for solving the Einstein's equation is to propose
an ansatz for the metric. That ansatz must be compatible with the
symmetries we would like to respect in the problem. The formal study
of the symmetries of the fields is accomplish via the Lie derivative
(for reviews, see
Refs. \cite{yano1957theory,Choquet:1989book,Nakahara,McInerney2013}). Below,
we use of the Lie derivative for obtaining ansätze for either the
metric or the connection. 

The form of the Lie derivative for tensors is well-known, but the Lie
derivative for a connection is not. Thus, for the sake of completeness
we remind the readers that
\begin{equation}
  \label{eq.Lie_Gamma}
  \mathcal{L}_{\xi} \Gamma^{a}{}_{bc} =
  \xi^m \partial_m \Gamma^{a}{}_{bc}
  - \Gamma^{m}{}_{bc} \partial_m \xi^a
  + \Gamma^{a}{}_{mc} \partial_b \xi^m
  + \Gamma^{a}{}_{bm} \partial_c \xi^m
  + \frac{\partial^2 \xi^{a}}{\partial x^{b} \partial x^c },
\end{equation}
where \(\xi\) is the vector defining the symmetry flow.

We shall restrict ourselves to the isotropic (spherically symmetric)
and, homogeneous and isotropic (cosmological symmetry). In the tedious
task of calculating the Lie derivative of different objects, we have
used the mathematical software SAGE together with its differential
geometry package SageManifolds \cite{sage,sagemanifolds}.

\subsection{Isotropic ansätze}
\label{sec.sphe_ans}
The two-dimensional sphere, \(S^2\), is a maximally symmetric space
whose Killing vectors generate an \(SO(3)\) symmetry group. The se
vectors can be expressed in spherical coordinates as
\begin{equation}
\begin{aligned}
J_1 &= 
\begin{pmatrix}
0 & 0 & -\cos\left({\varphi}\right) & \cot\left({\theta}\right)\sin\left({\varphi}\right)
\end{pmatrix},
\\
J_2 &= 
\begin{pmatrix}
0 & 0 & \sin\left({\varphi}\right) & \cos\left({\varphi}\right)\cot\left({\theta}\right)
\end{pmatrix},
\\
J_3 &= 
\begin{pmatrix}
0 & 0 & 0 & 1
\end{pmatrix}.
\end{aligned}
\label{Jab-sph}
\end{equation}

\subsubsection{Isotropic (covariant) two-tensor}
\label{sec.sph-2-tensor}
Let us start by finding the most general isotropic, four-dimensional,
covariant rank two-tensor. We shall obtain a generalisation of the
famous ansatz for the Schwarzschild metric.

We start from a general rank two tensor, i.e., the sixteen components
of the tensor depend on all the coordinates. Then, the Lie derivative
of the metric along the vector \(J_3\) in Eq. \eqref{Jab-sph} yields
\begin{equation}
  \label{Kill_J3}
  \pounds_{_{J_3}} T_{\mu\nu} = \frac{ \partial T_{\mu\nu} }{ \partial \varphi },
\end{equation}
which vanishes only if none of the components of the metric depend on
the \(\varphi\) coordinate. 

The Lie derivative along the other two generators of the angular
momentum yield a nontrivial set of differential equations (not shown
here)
whose solution fixes a tensor of the form,
\begin{equation}
  \label{eq:iso-tensor}
  T = 
  \begin{pmatrix}
    T_{00}\left(t, r\right) & T_{01}\left(t, r\right) & 0 & 0 \\
    T_{10}\left(t, r\right) & T_{11}\left(t, r\right) & 0 & 0 \\
    0 & 0 & T_{22}\left(t, r\right) & H\left(t, r\right) \sin\left({\theta}\right) \\
    0 & 0 & -H\left(t, r\right) \sin\left({\theta}\right) & T_{22}\left(t, r\right) \sin^2\left({\theta}\right)
  \end{pmatrix}.
\end{equation}

This result was found by A. Papapetrou \cite{Papapetrou:1948}. Notice
that there are six functions of the coordinates \(t\) and \(r\), while the
\(\theta\) dependence is fixed by the symmetry. Two out of the six
functions vanish whenever one restrict to symmetric tensors, i.e.
\(g(X,Y) = g(Y,X)\), such as the metric tensor. The form of the
symmetric, covariant, rank two tensor is
\begin{equation}
  g = \begin{pmatrix}
      A(t,r) & B(t,r) & 0 & 0 \\
      B(t,r) & C(t,r) & 0 & 0 \\
      0 & 0 & D(t,r) & 0 \\
      0 & 0 & 0 & D(t,r) \sin^2 (\theta)
      \end{pmatrix},
  \label{sph-met}
\end{equation}
which, under a redefinition of the radial and temporal coordinates,
takes the standard form,
\begin{equation}
  \label{Sch-met}
  g = \begin{pmatrix}
      F(t,r) & 0 & 0 & 0 \\
      0 & G(t,r) & 0 & 0 \\
      0 & 0 & D(t,r) & 0 \\
      0 & 0 & 0 & D(t,r) \sin^2 (\theta)
      \end{pmatrix}.
\end{equation}

It is worth noticing that Eq. \eqref{Sch-met} is the most general
spherical ansatz is one wants to solve Einstein's field equations. The
static condition is only assured by the Birkhoff theorem
\cite{Jebsen1921,Birkhoff1923,Alexandrow1923,Eisland1925}, once the
field equations are given.

\subsubsection{Isotropic affine connection}
\label{sec.sph-con}
The strategy used in the previous section can be repeated for an
affine connection, and we shall end up with the most general isotropic
(affine) connection. For the sake of simplicity, we do not include the
differential equations obtained from the calculation of the Lie
derivative.\footnote{Notice that in four dimensions and affine connection has
sixty-four components. Therefore, there are one hundred and ninety two
differential equations to solve for the three generators of spherical
symmetry.}

As before, the Lie derivative along the generators of the spherical
symmetry fix the angular dependence of the connection's
components. The nonvanishing components of an isotropic
\(\hat{\Gamma}^a{}_{bc}\), are
\begin{align*}
\hat{\Gamma}_{ \phantom{\, t } \, t \, t }^{ \, t \phantom{\, t
} \phantom{\, t } } & = F_{000}\left(t, r\right) 
&
 \hat{\Gamma}_{
\phantom{\, t } \, t \, r }^{ \, t \phantom{\, t } \phantom{\, r } }
& = F_{001}\left(t, r\right) 
\\
 \hat{\Gamma}_{ \phantom{\, t } \, r \, t
}^{ \, t \phantom{\, r } \phantom{\, t } } & = F_{010}\left(t,
r\right) 
&
 \hat{\Gamma}_{ \phantom{\, t } \, r \, r }^{ \, t \phantom{\, r }
\phantom{\, r } } & = F_{011}\left(t, r\right) 
\\
 \hat{\Gamma}_{
\phantom{\, t } \, {\theta} \, {\theta} }^{ \, t \phantom{\, {\theta} }
\phantom{\, {\theta} } } & = F_{033}\left(t, r\right) 
&
 \hat{\Gamma}_{
\phantom{\, t } \, {\theta} \, {\phi} }^{ \, t \phantom{\, {\theta} }
\phantom{\, {\phi} } } & = F_{023}\left(t, r\right)
\sin\left({\theta}\right) 
\\
 \hat{\Gamma}_{ \phantom{\, t } \, {\phi} \, {\theta}
}^{ \, t \phantom{\, {\phi} } \phantom{\, {\theta} } } & =
-F_{023}\left(t, r\right) \sin\left({\theta}\right) 
&
 \hat{\Gamma}_{ \phantom{\, t
} \, {\phi} \, {\phi} }^{ \, t \phantom{\, {\phi} } \phantom{\, {\phi} }
} & = F_{033}\left(t, r\right) \sin^2\left({\theta}\right)
\\
 \hat{\Gamma}_{ \phantom{\, r } \, t \, t }^{ \, r \phantom{\, t } \phantom{\, t
} } & = F_{100}\left(t, r\right) 
&
 \hat{\Gamma}_{ \phantom{\, r } \, t
\, r }^{ \, r \phantom{\, t } \phantom{\, r } } & =
F_{101}\left(t, r\right) 
\\
 \hat{\Gamma}_{ \phantom{\, r } \, r \, t }^{ \, r
\phantom{\, r } \phantom{\, t } } & = F_{110}\left(t, r\right)
&
 \hat{\Gamma}_{ \phantom{\, r } \, r \, r }^{ \, r \phantom{\, r } \phantom{\, r
} } & = F_{111}\left(t, r\right) 
\\
 \hat{\Gamma}_{ \phantom{\, r } \,
{\theta} \, {\theta} }^{ \, r \phantom{\, {\theta} } \phantom{\,
{\theta} } } & = F_{133}\left(t, r\right) 
&
 \hat{\Gamma}_{ \phantom{\, r
} \, {\theta} \, {\phi} }^{ \, r \phantom{\, {\theta} } \phantom{\,
{\phi} } } & = F_{123}\left(t, r\right)
\sin\left({\theta}\right) 
\\
 \hat{\Gamma}_{ \phantom{\, r } \, {\phi} \, {\theta}
}^{ \, r \phantom{\, {\phi} } \phantom{\, {\theta} } } & =
-F_{123}\left(t, r\right) \sin\left({\theta}\right) 
&
 \hat{\Gamma}_{ \phantom{\, r
} \, {\phi} \, {\phi} }^{ \, r \phantom{\, {\phi} } \phantom{\, {\phi} }
} & = F_{133}\left(t, r\right) \sin^2\left({\theta}\right)
\\
 \hat{\Gamma}_{ \phantom{\, {\theta} } \, t \, {\theta} }^{ \, {\theta}
\phantom{\, t } \phantom{\, {\theta} } } & = F_{303}\left(t,
r\right) 
&
 \hat{\Gamma}_{ \phantom{\, {\theta} } \, t \, {\phi} }^{ \, {\theta}
\phantom{\, t } \phantom{\, {\phi} } } & = -F_{302}\left(t,
r\right) \sin\left({\theta}\right) 
\\
 \hat{\Gamma}_{ \phantom{\, {\theta} } \, r \,
{\theta} }^{ \, {\theta} \phantom{\, r } \phantom{\, {\theta} } } & = F_{313}\left(t, r\right) 
&
 \hat{\Gamma}_{ \phantom{\, {\theta} } \, r \,
{\phi} }^{ \, {\theta} \phantom{\, r } \phantom{\, {\phi} } } & = -F_{312}\left(t, r\right) \sin\left({\theta}\right) 
\\
 \hat{\Gamma}_{
\phantom{\, {\theta} } \, {\theta} \, t }^{ \, {\theta} \phantom{\,
{\theta} } \phantom{\, t } } & = F_{330}\left(t, r\right) 
&
\hat{\Gamma}_{ \phantom{\, {\theta} } \, {\theta} \, r }^{ \, {\theta} \phantom{\,
{\theta} } \phantom{\, r } } & = F_{331}\left(t, r\right) 
\\
\hat{\Gamma}_{ \phantom{\, {\theta} } \, {\phi} \, t }^{ \, {\theta} \phantom{\,
{\phi} } \phantom{\, t } } & = -F_{320}\left(t, r\right)
\sin\left({\theta}\right) 
&
 \hat{\Gamma}_{ \phantom{\, {\theta} } \, {\phi} \, r
}^{ \, {\theta} \phantom{\, {\phi} } \phantom{\, r } } & =
-F_{321}\left(t, r\right) \sin\left({\theta}\right) 
\\
 \hat{\Gamma}_{ \phantom{\,
{\theta} } \, {\phi} \, {\phi} }^{ \, {\theta} \phantom{\, {\phi} }
\phantom{\, {\phi} } } & = -\cos\left({\theta}\right)
\sin\left({\theta}\right) 
&
 \hat{\Gamma}_{ \phantom{\, {\phi} } \, t \, {\theta}
}^{ \, {\phi} \phantom{\, t } \phantom{\, {\theta} } } & =
\frac{F_{302}\left(t, r\right)}{\sin\left({\theta}\right)} 
\\
 \hat{\Gamma}_{
\phantom{\, {\phi} } \, t \, {\phi} }^{ \, {\phi} \phantom{\, t }
\phantom{\, {\phi} } } & = F_{303}\left(t, r\right) 
&
 \hat{\Gamma}_{
\phantom{\, {\phi} } \, r \, {\theta} }^{ \, {\phi} \phantom{\, r }
\phantom{\, {\theta} } } & = \frac{F_{312}\left(t,
r\right)}{\sin\left({\theta}\right)} 
\\
 \hat{\Gamma}_{ \phantom{\, {\phi} } \, r \,
{\phi} }^{ \, {\phi} \phantom{\, r } \phantom{\, {\phi} } } & = F_{313}\left(t, r\right) 
&
 \hat{\Gamma}_{ \phantom{\, {\phi} } \, {\theta}
\, t }^{ \, {\phi} \phantom{\, {\theta} } \phantom{\, t } } & = \frac{F_{320}\left(t, r\right)}{\sin\left({\theta}\right)} 
\\
 \hat{\Gamma}_{
\phantom{\, {\phi} } \, {\theta} \, r }^{ \, {\phi} \phantom{\, {\theta}
} \phantom{\, r } } & = \frac{F_{321}\left(t,
r\right)}{\sin\left({\theta}\right)} 
&
 \hat{\Gamma}_{ \phantom{\, {\phi} } \,
{\theta} \, {\phi} }^{ \, {\phi} \phantom{\, {\theta} } \phantom{\,
{\phi} } } & =
\frac{\cos\left({\theta}\right)}{\sin\left({\theta}\right)} 
\\
 \hat{\Gamma}_{
\phantom{\, {\phi} } \, {\phi} \, t }^{ \, {\phi} \phantom{\, {\phi} }
\phantom{\, t } } & = F_{330}\left(t, r\right) 
&
 \hat{\Gamma}_{
\phantom{\, {\phi} } \, {\phi} \, r }^{ \, {\phi} \phantom{\, {\phi} }
\phantom{\, r } } & = F_{331}\left(t, r\right) 
\\
 \hat{\Gamma}_{
\phantom{\, {\phi} } \, {\phi} \, {\theta} }^{ \, {\phi} \phantom{\,
{\phi} } \phantom{\, {\theta} } } & =
\frac{\cos\left({\theta}\right)}{\sin\left({\theta}\right)}.
\end{align*}
This general isotropic connection, depends on twenty functions of the
coordinates \(t\) and \(r\), and has nonvanishing torsion and
non-metricity. Therefore, for our purposes within this paper, we can
restrict even further to a torsion-free connection, whose components
are
\begin{equation}
  \begin{aligned}
    \Gamma_{ \phantom{t } t t }^{ t
      \phantom{t } \phantom{t } } & = F_{000}\left(t, r\right)
    &
    \Gamma_{ \phantom{t } t r }^{ t \phantom{t }
      \phantom{r } } & = F_{001}\left(t, r\right)
    \\ 
    \Gamma_{ \phantom{t } r
      r }^{ t \phantom{r } \phantom{r } } & =
    F_{011}\left(t, r\right)
    &
    \Gamma_{ \phantom{t } {\theta} \,
      {\theta} }^{ t \phantom{{\theta} } \phantom{{\theta} } } & =
    F_{033}\left(t, r\right)
    \\ 
    \Gamma_{ \phantom{t } {\phi}
      {\phi} }^{ t \phantom{{\phi} } \phantom{{\phi} } } & =
    F_{033}\left(t, r\right) \sin^2\left({\theta}\right)
    &
    \Gamma_{
      \phantom{r } t t }^{ r \phantom{t } \phantom{t } }
    & = F_{100}\left(t, r\right)
    \\ 
    \Gamma_{ \phantom{r } t
      r }^{ r \phantom{t } \phantom{r } } & =
    F_{101}\left(t, r\right)
    &
    \Gamma_{ \phantom{r } r r }^{ r \phantom{r }
      \phantom{r } } & = F_{111}\left(t, r\right)
    \\
    \Gamma_{
      \phantom{r } {\theta} {\theta} }^{ r \phantom{{\theta} }
      \phantom{{\theta} } } & = F_{133}\left(t, r\right)
    & 
    \Gamma_{ \phantom{r } {\phi} {\phi} }^{ r \phantom{{\phi}
      } \phantom{{\phi} } } & = F_{133}\left(t, r\right)
    \sin^2\left({\theta}\right)
    \\
    \Gamma_{ \phantom{{\theta} } t \,
      {\theta} }^{ {\theta} \phantom{t } \phantom{{\theta} } } & =
    F_{303}\left(t, r\right)
    & 
    \Gamma_{ \phantom{{\theta} } t
      {\phi} }^{ {\theta} \phantom{t } \phantom{{\phi} } } & =
    -F_{302}\left(t, r\right) \sin\left({\theta}\right)
    \\
    \Gamma_{
      \phantom{{\theta} } r {\theta} }^{ {\theta} \phantom{r }
      \phantom{{\theta} } } & = F_{313}\left(t, r\right)
    & 
    \Gamma_{ \phantom{{\theta} } r {\phi} }^{ {\theta}
      \phantom{r } \phantom{{\phi} } } & = -F_{312}\left(t,
    r\right) \sin\left({\theta}\right)
    \\
    \Gamma_{
      \phantom{{\theta} } {\phi} {\phi} }^{ {\theta} \phantom{\,
        {\phi} } \phantom{{\phi} } } & = -\cos\left({\theta}\right)
    \sin\left({\theta}\right)
    & 
    \Gamma_{ \phantom{{\phi} } t \,
      {\theta} }^{ {\phi} \phantom{t } \phantom{{\theta} } } & =
    \frac{F_{302}\left(t, r\right)}{\sin\left({\theta}\right)}
    \\
    \Gamma_{ \phantom{{\phi} } t {\phi} }^{ {\phi} \phantom{t
      } \phantom{{\phi} } } & = F_{303}\left(t, r\right)
    & 
    \Gamma_{ \phantom{{\phi} } r {\theta} }^{ {\phi} \phantom{\,
        r } \phantom{{\theta} } } & = \frac{F_{312}\left(t,
      r\right)}{\sin\left({\theta}\right)}
    \\
    \Gamma_{ \phantom{{\phi} } \,
      r {\phi} }^{ {\phi} \phantom{r } \phantom{{\phi} } } & =
    F_{313}\left(t, r\right)
    & 
    \Gamma_{
      \phantom{{\phi} } {\theta} {\phi} }^{ {\phi} \phantom{\,
        {\theta} } \phantom{{\phi} } } & =
    \frac{\cos\left({\theta}\right)}{\sin\left({\theta}\right)}.
  \end{aligned}
  \label{eq:isot-conn-torsion-free}
\end{equation}
The last connection depends on twelve functions of \(t\) and \(r\), and
these are the functions to be fixed by solving the field equations
\eqref{SimpleEOM}.

Using the Eq. \eqref{eq:isot-conn-torsion-free}, we calculated the Ricci
tensor and noticed that it is not symmetric, because the connection is
not equi-affine. In order for the Ricci to be symmetric, the following
conditions must hold,
\begin{equation}
\label{eq:cond1-equiaffine}
\frac{ \partial F_{000} }{ \partial r} - \frac{ \partial F_{001} }{ \partial t} + 2 \frac{ \partial F_{303} }{ \partial r} = 0,
\end{equation}
and
\begin{equation}
\label{eq:cond2-equiaffine}
\frac{ \partial F_{101} }{ \partial r} - \frac{ \partial F_{101} }{ \partial t} + 2 \frac{ \partial F_{313} }{ \partial r} = 0.
\end{equation}
There are several solutions to these conditions, and each of them
could (in principle) provides a solution to the field equations
\eqref{SimpleEOM}. It is worth mentioning that if we do not demand the
connection to be equi-affine, the field equations to solve would be
Eq. \eqref{almostSimpleEOM}, which have and extra term which cannot be
obtained from the Yang--Mills-like effective action in Eq. \eqref{SKY}.

\subsection{Cosmological ansätze}
\label{sec.cosm_ans}
The Lorentian isotropic and homogeneous spaces in four dimensions have
isometry group either \(SO(4)\), \(SO(3,1)\) or \(ISO(3)\), and their
algebra can be obtained from the algebra \(\mathfrak{so}(4)\) through
a \(3+1\) decomposition, i.e. \(J_{AB} = \set{ J_{ab}, J_{a*} }\),\footnote{We shall use the standard (Euclidean) three-dimensional
correspondence \(J_a = \frac{1}{2} \epsilon_{abc} J^{bc}\), in order
to match notations with the previous case.}
where the extra dimension has been denoted by an asterisk. In term of
these new generators, the algebra reads
\begin{equation}
  \label{alg-SO-decomp}
  \begin{split}
    \comm{ J_{ab} }{ J_{cd} } &= \delta_{bc} J_{ad} - \delta_{ac} J_{bd} + \delta_{ad} J_{bc} - \delta_{bd} J_{ac}, \\
    \comm{ J_{ab} }{ J_{c*} } &= \delta_{bc} J_{a*} - \delta_{ac} J_{b*}, \\
    \comm{ J_{a*} }{ J_{c*} } &= - \kappa J_{ac},
  \end{split}
\end{equation} 
with\footnote{The inhomogeneous algebra of \(ISO(n)\) can be obtained from
the ones of \(SO(n+1)\) or \(SO(n,1)\) through the Inönü--Wigner
contraction \cite{Gilmore}.}
\begin{equation}
  \kappa = 
  \begin{cases}
    1  & SO(4) \\
    0  & ISO(3) \\
    -1 & SO(3,1)
  \end{cases}.
\end{equation}

One can express the Killing vectors in spherical coordinates, and in
addition to those obtained in Eq.  \eqref{Jab-sph} we get
\begin{equation}
  \begin{aligned}
  P_1 = J_{1*} &= \sqrt{1-{\kappa} r^{2}}
  \begin{pmatrix}
  0 &  \cos\left({\varphi}\right)
  \sin\left({\theta}\right) & 
  \frac{\cos\left({\varphi}\right) \cos\left({\theta}\right)}{r} &
  - \frac{\sin\left({\varphi}\right)}{r
  \sin\left({\theta}\right)}
  \end{pmatrix},
  \\
  P_2 = J_{2*} &= \sqrt{1-{\kappa} r^{2}}
  \begin{pmatrix}
  0 &  \sin\left({\varphi}\right)
  \sin\left({\theta}\right) & \frac{
  \cos\left({\theta}\right) \sin\left({\varphi}\right)}{r} &
  \frac{ \cos\left({\varphi}\right)}{r
  \sin\left({\theta}\right)}
  \end{pmatrix},
  \\
  P_3 = J_{3*} &= \sqrt{1-{\kappa} r^{2}}
  \begin{pmatrix}
  0 &  \cos\left({\theta}\right) &
  -\frac{ \sin\left({\theta}\right)}{r} &
  0
  \end{pmatrix}.
  \end{aligned}
  \label{Kill-homotropic}
\end{equation}

\subsubsection{Isotropic and homogeneous (covariant) two-tensor}
\label{sec.homotropic-tensor}
In order to find the most general isotropic and homogeneous,
covariant two-tensor, we can start from the result in
Eq. \eqref{eq:iso-tensor}, and impose now the symmetries from the extra
generators. The equations are\footnote{Due to the isotropy, one needs to solve just for one of these
extra generators.}
\begin{equation}
  \begin{aligned}
    \pounds_{_{P_{3}}} T_{  t  t }^{ \phantom{ t } \phantom{ t } } & : & \sqrt{ 1 - {\kappa} r^{2} } \cos\left({\theta}\right) \frac{\partial T_{00}}{\partial r} & = 0,
    &
    \pounds_{_{P_{3}}} T_{  t  r }^{ \phantom{ t } \phantom{ r } } & : & {\left({\kappa} r T_{01}  + {\left({\kappa} r^{2} - 1\right)} \frac{\partial T_{01}}{\partial r}\right)} & = 0,
    \\
    \pounds_{_{P_{3}}} T_{  t  {\theta} }^{ \phantom{ t } \phantom{ {\theta} } } & : &  T_{01} & = 0,
    &
    \pounds_{_{P_{3}}} T_{  r  t }^{ \phantom{ r } \phantom{ t } } & : & {\left({\kappa} r T_{10}  + {\left({\kappa} r^{2} - 1\right)} \frac{\partial T_{10}}{\partial r}\right)} & = 0,
    \\
    \pounds_{_{P_{3}}} T_{  r  r }^{ \phantom{ r } \phantom{ r } } & : &  {\left(2  {\kappa} r T_{11}  + {\left({\kappa} r^{2} - 1\right)} \frac{\partial T_{11}}{\partial r}\right)} & = 0,
    &
    \pounds_{_{P_{3}}} T_{  r  {\theta} }^{ \phantom{ r } \phantom{ {\theta} } } & : & {\left({\kappa} r^{4} - r^{2}\right)} T_{11}  + T_{22} & = 0,
    \\
    \pounds_{_{P_{3}}} T_{  r  {\varphi} }^{ \phantom{ r } \phantom{ {\varphi} } } & : & H & = 0,
    &
    \pounds_{_{P_{3}}} T_{  {\theta}  t }^{ \phantom{ {\theta} } \phantom{ t } } & : &  T_{10} & = 0,
    \\
    \pounds_{_{P_{3}}} T_{  {\theta}  r }^{ \phantom{ {\theta} } \phantom{ r } } & : & \left({\kappa} r^{4} - r^{2}\right) T_{11}  + T_{22} & = 0,
    &
    \pounds_{_{P_{3}}} T_{  {\theta}  {\theta} }^{ \phantom{ {\theta} } \phantom{ {\theta} } } & : & {\left(r \frac{\partial T_{22}}{\partial r} - 2  T_{22} \right)} & = 0,
    \\
    \pounds_{_{P_{3}}} T_{  {\theta}  {\varphi} }^{ \phantom{ {\theta} } \phantom{ {\varphi} } } & : &  {\left(r \frac{\partial H}{\partial r} - 2  H \right)} & = 0,
    &
    \pounds_{_{P_{3}}} T_{  {\varphi}  r }^{ \phantom{ {\varphi} } \phantom{ r } } & : & H & = 0,
    \\
    \pounds_{_{P_{3}}} T_{  {\varphi}  {\theta} }^{ \phantom{ {\varphi} } \phantom{ {\theta} } } & : & {\left(r \frac{\partial H}{\partial r} - 2  H \right)} & = 0,
    &
    \pounds_{_{P_{3}}} T_{  {\varphi}  {\varphi} }^{ \phantom{ {\varphi} } \phantom{ {\varphi} } } & : & {\left(r \frac{\partial T_{22}}{\partial r} - 2  T_{22} \right)} & = 0.
  \end{aligned}
\end{equation}

The above equations are solved for a tensor of the form,
\begin{equation}
  T = G_{00}\left(t\right) \mathrm{d} t\otimes \mathrm{d} t + \frac{G_{11}\left(t\right)}{ 1 - {\kappa} r^{2} } \mathrm{d}
  r\otimes \mathrm{d} r + r^{2} G_{11}\left(t\right) \mathrm{d}
  {\theta}\otimes \mathrm{d} {\theta} + r^{2} G_{11}\left(t\right)
  \sin^2\left({\theta}\right) \mathrm{d} {\varphi}\otimes \mathrm{d} {\varphi}.
\end{equation}
Under a redefinition of the ``time'' coordinate, the tensor is nothing
but the standard ansatz for the Friedman--Robertson--Walker
metric. 

\subsubsection{Isotropic And homogeneous affine connection}
\label{sec.homotropic-connection}
Without further details, we present the nonvanishing components of an
isotropic and homogeneous affine connection. It is given by,
\begin{equation}
  \begin{aligned}
    \hat{\Gamma}_{ \phantom{  t }   t   t }^{   t \phantom{  t } \phantom{  t } } & = G_{000}\left(t\right) &   \hat{\Gamma}_{ \phantom{  t }   r   r }^{   t \phantom{  r } \phantom{  r } } & = \frac{G_{011}\left(t\right)}{ 1 - {\kappa} r^{2} }
    \\
    \hat{\Gamma}_{ \phantom{  t }   {\theta}   {\theta} }^{   t \phantom{  {\theta} } \phantom{  {\theta} } } & = r^{2} G_{011}\left(t\right) &  \hat{\Gamma}_{ \phantom{  t }   {\varphi}   {\varphi} }^{   t \phantom{  {\varphi} } \phantom{  {\varphi} } } & = r^{2} G_{011}\left(t\right) \sin^2\left({\theta}\right)
    \\
    \hat{\Gamma}_{ \phantom{  r }   t   r }^{   r \phantom{  t } \phantom{  r } } & = G_{101}\left(t\right) &  \hat{\Gamma}_{ \phantom{  r }   r   t }^{   r \phantom{  r } \phantom{  t } } & = G_{110}\left(t\right)
    \\
    \hat{\Gamma}_{ \phantom{  r }   r   r }^{   r \phantom{  r } \phantom{  r } } & = \frac{{\kappa} r}{ 1 - {\kappa} r^{2} } &   \hat{\Gamma}_{ \phantom{  r }   {\theta}   {\theta} }^{   r \phantom{  {\theta} } \phantom{  {\theta} } } & = {\kappa} r^{3} - r
    \\
    \hat{\Gamma}_{ \phantom{  r }   {\theta}   {\varphi} }^{   r \phantom{  {\theta} } \phantom{  {\varphi} } } & = \sqrt{1-{\kappa} r^{2}} r^{2} G_{123}\left(t\right) \sin\left({\theta}\right) &   \hat{\Gamma}_{ \phantom{  r }   {\varphi}   {\theta} }^{   r \phantom{  {\varphi} } \phantom{  {\theta} } } & = -\sqrt{1-{\kappa} r^{2}} r^{2} G_{123}\left(t\right) \sin\left({\theta}\right)
    \\
    \hat{\Gamma}_{ \phantom{  r }   {\varphi}   {\varphi} }^{   r \phantom{  {\varphi} } \phantom{  {\varphi} } } & = {\left({\kappa} r^{3} - r\right)} \sin^2\left({\theta}\right) &   \hat{\Gamma}_{ \phantom{  {\theta} }   t   {\theta} }^{   {\theta} \phantom{  t } \phantom{  {\theta} } } & = G_{101}\left(t\right)
    \\
    \hat{\Gamma}_{ \phantom{  {\theta} }   r   {\theta} }^{   {\theta} \phantom{  r } \phantom{  {\theta} } } & = \frac{1}{r} &   \hat{\Gamma}_{ \phantom{  {\theta} }   r   {\varphi} }^{   {\theta} \phantom{  r } \phantom{  {\varphi} } } & = -\frac{G_{123}\left(t\right) \sin\left({\theta}\right)}{\sqrt{1-{\kappa} r^{2}}}
    \\
    \hat{\Gamma}_{ \phantom{  {\theta} }   {\theta}   t }^{   {\theta} \phantom{  {\theta} } \phantom{  t } } & = G_{110}\left(t\right) &   \hat{\Gamma}_{ \phantom{  {\theta} }   {\theta}   r }^{   {\theta} \phantom{  {\theta} } \phantom{  r } } & = \frac{1}{r}
    \\
    \hat{\Gamma}_{ \phantom{  {\theta} }   {\varphi}   r }^{   {\theta} \phantom{  {\varphi} } \phantom{  r } } & = \frac{G_{123}\left(t\right) \sin\left({\theta}\right)}{\sqrt{1-{\kappa} r^{2}}} &   \hat{\Gamma}_{ \phantom{  {\theta} }   {\varphi}   {\varphi} }^{   {\theta} \phantom{  {\varphi} } \phantom{  {\varphi} } } & = -\cos\left({\theta}\right) \sin\left({\theta}\right)
    \\
    \hat{\Gamma}_{ \phantom{  {\varphi} }   t   {\varphi} }^{   {\varphi} \phantom{  t } \phantom{  {\varphi} } } & = G_{101}\left(t\right) &   \hat{\Gamma}_{ \phantom{  {\varphi} }   r   {\theta} }^{   {\varphi} \phantom{  r } \phantom{  {\theta} } } & = \frac{G_{123}\left(t\right)}{\sqrt{1-{\kappa} r^{2}} \sin\left({\theta}\right)}
    \\
    \hat{\Gamma}_{ \phantom{  {\varphi} }   r   {\varphi} }^{   {\varphi} \phantom{  r } \phantom{  {\varphi} } } & = \frac{1}{r} &   \hat{\Gamma}_{ \phantom{  {\varphi} }   {\theta}   r }^{   {\varphi} \phantom{  {\theta} } \phantom{  r } } & = -\frac{G_{123}\left(t\right)}{\sqrt{1-{\kappa} r^{2}} \sin\left({\theta}\right)}
    \\
    \hat{\Gamma}_{ \phantom{  {\varphi} }   {\theta}   {\varphi} }^{   {\varphi} \phantom{  {\theta} } \phantom{  {\varphi} } } & =  \frac{\cos\left({\theta}\right)}{\sin\left({\theta}\right)} &  \hat{\Gamma}_{ \phantom{  {\varphi} }   {\varphi}   t }^{   {\varphi} \phantom{  {\varphi} } \phantom{  t } } & = G_{110}\left(t\right)
    \\
    \hat{\Gamma}_{ \phantom{  {\varphi} }   {\varphi}   r }^{   {\varphi} \phantom{  {\varphi} } \phantom{  r } } & = \frac{1}{r} &   \hat{\Gamma}_{ \phantom{  {\varphi} }   {\varphi}   {\theta} }^{   {\varphi} \phantom{  {\varphi} } \phantom{  {\theta} } } & = \frac{\cos\left({\theta}\right)}{\sin\left({\theta}\right)},
  \end{aligned}
\end{equation}
which is determined by five independent functions, but this affine
connection still possess torsion. The imposition of vanishing torsion
kills two of the above functions, and the remaining components of the
connection are
\begin{equation}
  \label{eq.homotropic_conn_nt}
  \begin{aligned}
    \Gamma_{ \phantom{  t }   t   t }^{   t \phantom{  t } \phantom{  t } }
    & = G_{000}\left(t\right)
    &
    \Gamma_{ \phantom{  t }   r   r }^{   t \phantom{  r } \phantom{  r } } & = \frac{G_{011}\left(t\right)}{ 1 - {\kappa} r^{2} } 
    \\
    \Gamma_{ \phantom{  t }   {\theta}   {\theta} }^{   t \phantom{  {\theta} } \phantom{
        {\theta} } } & = r^{2} G_{011}\left(t\right)
    &
    \Gamma_{ \phantom{  t }   {\varphi}   {\varphi} }^{   t \phantom{  {\varphi} } \phantom{  {\varphi} } } & = r^{2} G_{011}\left(t\right) \sin^2\left({\theta}\right) 
    \\
    \Gamma_{ \phantom{  r }   t   r }^{   r \phantom{  t } \phantom{  r } }
    & = G_{101}\left(t\right)
    &
    \Gamma_{ \phantom{  r }   r   r }^{   r \phantom{  r } \phantom{  r } }
    & = \frac{{\kappa} r}{ 1 - {\kappa} r^{2} }
    \\
    \Gamma_{ \phantom{  r }   {\theta}   {\theta} }^{   r \phantom{  {\theta} } \phantom{  {\theta} } } & = {\kappa} r^{3} - r 
    &
    \Gamma_{ \phantom{  r }   {\varphi}   {\varphi} }^{   r \phantom{
        {\varphi} } \phantom{  {\varphi} } } & = {\left({\kappa} r^{3} -
      r\right)} \sin^2\left({\theta}\right)
    \\
    \Gamma_{ \phantom{  {\theta} }   t   {\theta} }^{   {\theta} \phantom{  t } \phantom{  {\theta} } } & = G_{101}\left(t\right) 
    &
    \Gamma_{ \phantom{  {\theta} }   r   {\theta} }^{   {\theta} \phantom{  r } \phantom{
        {\theta} } } & = \frac{1}{r}
    \\
    \Gamma_{ \phantom{  {\theta} }   {\varphi}   {\varphi} }^{   {\theta}
      \phantom{  {\varphi} } \phantom{  {\varphi} } } & =
    -\cos\left({\theta}\right) \sin\left({\theta}\right) 
    &
    \Gamma_{ \phantom{  {\varphi} }   t   {\varphi} }^{   {\varphi}
      \phantom{  t } \phantom{  {\varphi} } } & =
    G_{101}\left(t\right)
    \\
    \Gamma_{ \phantom{  {\varphi} }   r   {\varphi} }^{   {\varphi}
      \phantom{  r } \phantom{  {\varphi} } } & = \frac{1}{r} 
    &
    \Gamma_{ \phantom{  {\varphi} }   {\theta}   {\varphi} }^{   {\varphi}
      \phantom{  {\theta} } \phantom{  {\varphi} } } & =
    \frac{\cos\left({\theta}\right)}{\sin\left({\theta}\right)}.
  \end{aligned}
\end{equation}

\section{Toward the solution of the field equations}
\label{sec.PAG_sol}
\subsection{Cosmological solutions}
\label{sec.PAG_sol_cosm}
Using the connection in Eq. \eqref{eq.homotropic_conn_nt}, the Ricci is calculated, and yields
\begin{equation}
  \begin{aligned}
    R_{  t  t }^{ \phantom{ t } \phantom{ t } }
    & =  3  G_{000}\left(t\right) G_{101}\left(t\right) - 3 
    G_{101}\left(t\right)^{2} - 3  \frac{\partial G_{101}}{\partial t} 
    \\
    R_{  r  r }^{ \phantom{ r } \phantom{ r } } & = 
    \frac{G_{000}\left(t\right) G_{011}\left(t\right) +
      G_{011}\left(t\right) G_{101}\left(t\right) + 2  {\kappa} +
      \frac{\partial G_{011}}{\partial t}}{1 - {\kappa} r^{2} } 
    \\
    R_{ 
      {\theta}  {\theta} }^{ \phantom{ {\theta} } \phantom{ {\theta} } }
    & =  \left( G_{000}\left(t\right) G_{011}\left(t\right) + 
    G_{011}\left(t\right) G_{101}\left(t\right) + 2  {\kappa}  +
    \frac{\partial G_{011}}{\partial t} \right) r^{2} 
    \\
    R_{  {\varphi}  {\varphi}
    }^{ \phantom{ {\varphi} } \phantom{ {\varphi} } } & = 
    {\left( G_{000}\left(t\right) G_{011}\left(t\right) + 
      G_{011}\left(t\right) G_{101}\left(t\right) + 2  {\kappa}  +
      \frac{\partial G_{011}}{\partial t}\right)} r^{2} 
    \sin^2\left({\theta}\right).
  \end{aligned}
\end{equation}
Notice that the Ricci tensor is symmetric, determined by only three
independent functions and, as expected by the symmetry, just two of
their components behave differently.

A first kind of solutions can be found by solving the system of
equations determined by vanishing Ricci. However, this strategy
requires the fixing of one of the unknown functions. A solution inspired in the
components of the connection for Friedmann--Robertson--Walker, gives
\begin{align}
  G_{000} &= 0,
  &
  G_{101} &= \frac{1}{t - C_1},
  &
  G_{011} &= - 2 \kappa t + C_2.
  \label{eq.sol_cosm_very_simple}
\end{align}

A second class of solutions can be found by solving the parallel Ricci
equation, \(\nabla_\lambda R_{\mu\nu} = 0\), which surprisingly yield
three independent field equations,
\begin{align}
  \nabla_t R_{tt} & = -6 \, G_{000}^{2} G_{101} + 6 \,
  G_{000} G_{101}^{2} + 3 \,
  G_{101} \frac{\partial\,G_{000}}{\partial t} + 3 \,
  {\left(3 \, G_{000} - 2 \, G_{101}\right)}
  \frac{\partial\,G_{101}}{\partial t} - 3 \,
  \frac{\partial^2\,G_{101}}{\partial t^2},
  \label{eq.DR_ttt}
  \\
  \nabla_t R_{ii} & \sim 2 \, G_{011} G_{101}^{2} + 2 \,
   {\left(G_{000} G_{011} + 2 \,
     {\kappa}\right)} G_{101} - G_{011}
   \frac{\partial\,G_{000}}{\partial t} ,
   \label{eq.DR_tii}
   \\ 
   & \quad - {\left(G_{000} -
     G_{101}\right)} \frac{\partial\,G_{011}}{\partial t} -
   G_{011} \frac{\partial\,G_{101}}{\partial t} -
   \frac{\partial^2\,G_{011}}{\partial t^2}
   \notag
   \\
   \nabla_i R_{t i} & = 2 \, G_{011} G_{101}^{2} - 2 \,
    {\left(2 \, G_{000} G_{011} +
      {\kappa}\right)} G_{101} - G_{101}
    \frac{\partial\,G_{011}}{\partial t} + 3 \, G_{011}
    \frac{\partial\,G_{101}}{\partial t}.
    \label{eq.DR_iti}
\end{align}
However, the system of equations is complicated enough to avoid an
analytic solution.\footnote{Of course one can propose a formal solution in terms of power
series. However, this approach is been used in a paper still under develop.} Despite the complication, we can try a couple
of assumptions that simplify the system of equations, for example:
\begin{itemize}
\item Again, inspired in the Friedmann--Robertson--Walker results we
choose \(G_{000}=0\), and solve for the other functions (see Ref. \cite{OCF-future2}).
\item We could choose \(G_{000} = G_{101}\), to eliminate the nonlinear
terms in Eq.  \eqref{eq.DR_ttt}.
\item We can solve both Eq. \eqref{ScalarAction} and  \eqref{eq.DR_iti} by
setting \(G_{011}= 0\), \(\kappa \neq 0\) and (say) \(G_{101} =
  \text{const.}\), and solve Eq. \eqref{eq.DR_ttt} which turns to be a
known ordinary differential equation.
\end{itemize}

Finally, the third class of solutions are those of
Eq. \eqref{SimpleEOM}. The set of equations degenerate and yield a
single independent field equation,
\begin{equation}
  {4 \, G_{011} G_{101}^{2} - 2 \,
    {\left(G_{000} G_{011} - {\kappa}\right)}
    G_{101} - G_{011}
    \frac{\partial\,G_{000}}{\partial t} - G_{000}
    \frac{\partial\,G_{011}}{\partial t} + 2 \, G_{011}
    \frac{\partial\,G_{101}}{\partial t} -
    \frac{\partial^2\,G_{011}}{\partial t^2}} = 0.
\end{equation}
Therefore, we need to set two out of the three unknown functions to be
able of solving for the connection.

\section{Conclusions}
\label{sec:org386b7a7}

We have presented a short review of the polynomial affine gravity,
whose field equations (in the torsion-free sector) generalises whose
of the standard general relativity. In the mentioned approximation,
the field equations coincide with (part of) those of a gravitational
Yang--Mills theory of gravity, known in the literature as the SKY
model.

Among the features of the polynimial affine gravity, we highlighted
the following:
\begin{itemize}
\item Although our spacetime could be metric, the metric plays no role in
the model building.
\item The non-relativistic limit of the model yields a Keplerian
potential, even with the contributions of torsion and non-metricity.
\item In the torsion-free sector, all the vacuum solutions to the Einstein
gravity are solutions of the polynomial affine gravity.
\item Scalar matter can be coupled to the polynomial affine gravity
through a symmetric, \(\binom{2}{0}\)-tensor density. The coupled
field equations can be written in a similar form to those of general
relativity, with the subtlety that where ever the metric appears in
the Einstein's equations, we just need a covariantly constant,
symmetric, \(\binom{0}{2}\)-tensor.
\item Although the model is built up without the necessity of a metric,
one can still assume that the connection is a metric potential. Such
consideration yields to obtain new solutions to the
Eq. \eqref{SimpleEOM}, which are not solutions of the standard
Einstein's equations.\footnote{Some solutions will be presented in Ref. \cite{OCF-future2}.}
\item We found the general ansatz for the connection, compatible with
isotropic and cosmological symmetries. Additionally, we have
sketched the road to solving the Eq. \eqref{SimpleEOM}, for the
cosmological connection ansatz. A thorough analysis will be
presented in Ref. \cite{OCF-future2}.
\end{itemize}

We would like to finish commenting about the necessity of coupling other
form of matters and a formal counting of degrees of freedoms. 

\subsection*{Acknowledgements}
\label{sec:org9314eaf}
O.C-F. wants to thank to the ICTP-SAIFR and IFT-UNESP (Sao
Paulo-Brasil) for the hospitality while finishing this article. This
work as been partially founded by the CONICYT (Chile) project
PAI-79140040. The
Centro Cient\'ifico Tecnol\'ogico de Valpara\'iso
(CCTVal) is funded by the Chilean Government through the Centers of
Excellence Basal Financing Program FB0821 of CONICYT.


\providecommand{\href}[2]{#2}\begingroup\raggedright\endgroup

\end{document}